\documentclass[aps,prl,twocolumn,floatfix,altaffilletter,superscriptaddress,preprintnumbers, tightenlines,showpacs,showkeys,nofootinbib,notoccite]{revtex4-1}
\usepackage[utf8]{inputenc}
\usepackage[colorlinks=true,citecolor=blue,linkcolor=blue]{hyperref}
\usepackage[normalem]{ulem}
\usepackage{amsmath,amssymb, mathrsfs,esvect}
\usepackage{epsfig}
\usepackage{graphicx}               
\usepackage{url}
\usepackage{color}
\usepackage{slashed}
\usepackage{multirow}
\usepackage{placeins}
\usepackage[dvipsnames]{xcolor}
\usepackage{epstopdf}
\usepackage{soul}
\usepackage{tikz}
\usepackage[capitalise, english]{cleveref}
\usepackage{siunitx}
\usepackage{xspace}
\usepackage{booktabs}
\usetikzlibrary{trees}
\usetikzlibrary{decorations.pathmorphing}
\usetikzlibrary{decorations.markings}
\newcommand\myshade{80}
\colorlet{mylinkcolor}{Blue}
\colorlet{mycitecolor}{Red}
\colorlet{myurlcolor}{violet}

\newcommand{\beq}{\begin{eqnarray}}
\newcommand{\eeq}{\end{eqnarray}}

\hypersetup{
	linkcolor  = mylinkcolor!\myshade!black,
	citecolor  = mycitecolor!\myshade!black,
	urlcolor   = myurlcolor!\myshade!black,
	colorlinks = true
}
\allowdisplaybreaks
\begin{document}

\title{A Continuous Galactic Line Source of Axions:  The Remarkable Case of $^{23}$Na}

\author{W. C. Haxton}
\email{haxton@berkeley.edu}
\affiliation{Department of Physics, University of California, Berkeley, CA 94720, USA}
\affiliation{Institute for Nuclear Theory, University of Washington, Seattle, WA 98195, USA}
\affiliation{Lawrence Berkeley National Laboratory, Berkeley, CA 94720, USA}
\author{Xing Liu} 
\email{xingyzt@berkeley.edu}
\affiliation{Department of Physics, University of California, Berkeley, CA 94720, USA}
\author{Annie McCutcheon} 
\email{amccutcheon@ucdavis.edu}
\affiliation{Department of Physics, University of California, Davis, CA 95616, USA}
\author{Anupam Ray}
\email{anupam.ray@berkeley.edu}
\affiliation{Department of Physics, University of California, Berkeley, CA 94720, USA}

\date{\today}
\begin{abstract}
We argue that $^{23}$Na is a potentially significant source of galactic axions. For temperatures $\gtrsim 7 \times 10^8$K -- characteristic of carbon burning in the massive progenitors of supernovae and ONeMg white dwarfs -- the 440 keV first excited state of $^{23}$Na is thermally  populated, with its repeated decays pumping stellar energy into escaping axions. Odd-A nuclear abundances are typically very low in high-temperature stellar environments (or absent entirely due to burn-up). $^{23}$Na is an exception: $\approx 0.1 M_\odot$ of the isotope is synthesized during carbon burning then maintained at $\approx 10^9$K for times ranging up to $6 \times 10^4$y.  Using  MESA simulations, a galactic model, and sampling over progenitor masses, locations, and evolutionary stages, we find a continuous flux at Earth of $\langle \phi_a \rangle \approx 22$/cm$^2$s for $g^\mathrm{eff}_{aNN} = 10^{-9}$.  Some fraction of these axions converts to photons as they propagate through the galactic magnetic field, producing a distinctive 440 keV line $\gamma$-ray detectable by all-sky detectors like the Compton Spectrometer and Imager (COSI). Assuming a 1$\mu$G galactic magnetic field and a sufficiently light axion mass, we find that COSI will be able to probe $| g_{aNN}^\mathrm{eff} g_{a \gamma \gamma} | \gtrsim1.8 \times 10^{-22}$ GeV$^{-1}$ at $3\sigma$ after two years of surveying.
\end{abstract}

\maketitle
\preprint{N3AS-24-014, LA-UR-24-24058}
In 1977 Peccei and Quinn proposed a solution \cite{PhysRevLett.38.1440} to the strong CP problem that depended on the introduction of a new scalar field.  Soon after, Weinberg  \cite{PhysRevLett.40.223} and Wilczek \cite{PhysRevLett.40.279} independently observed that a consequence of the Peccei--Quinn mechanism would be a new particle, the axion. Currently the axion is one of the most popular candidates for the dark matter, albeit characterized by couplings to standard-model particles considerably weaker than those originally envisioned.

Many of the existing constraints on axion-matter couplings depend on astrophysical sources. A celebrated example is the early universe, where axions are produced from the harmonic oscillations of the axion potential as it seeks its minimum, damping that motion. Sikivie \cite{PhysRevLett.51.1415,RevModPhys.93.015004} proposed that cosmological axions could be observed
through their conversion to microwave photons of the same energy.  ADMX is searching for these conversion photons, using a magnetic cavity tuned to absorb the missing momentum \cite{PhysRevD.64.092003,PhysRevLett.120.151301,PhysRevLett.124.101303,PhysRevLett.127.261803}. Various related experiments are underway~\cite{PhysRevD.97.092001,PhysRevD.106.103008,PhysRevLett.127.081801,PhysRevX.4.021030,quiskamp2024nearquantumlimitedaxiondark}. Axions or axion-like particles (ALPs) are a potential source of X-rays from both young \cite{PhysRevLett.106.081101,10.1111/j.1745-3933.2011.01015.x,LEINSON201587,Leinson_2014,PhysRevD.98.103015,Leinson}
and older \cite{PhysRevLett.128.091102,PhysRevD.93.065044,PhysRevD.99.043011} neutron stars. ALPs can alter the cooling of various stars \cite{Caputo:2024oqc,Carenza:2024ehj}, 
including transient events like supernovae \cite{Raffelt:2006cw,PhysRevD.94.085012,ChangJH,Carenza,PhysRevLett.126.071102}.

An interesting astrophysical source was pointed out by Haxton and Lee \cite{PhysRevLett.66.2557}: certain nuclei with low-lying M1 transitions can serve as axion pumps, converting stellar thermal energy into axions through repeated photo-excitation and axio-deexcitation.
They studied this mechanism as a function of stellar core temperature $(T)$, identifying three isotopes of special importance.  At low temperatures $T_8 = T/10^8\,\rm{K} \approx 1$, {$^{57}$Fe} (2.2\%  of natural Fe) dominates due to its 14.4 keV excited state, made famous by Mossbauer spectroscopy.  As this temperature is characteristic of red giants, they used such stars to probe the hadronic axion ``Turner window"~\cite{TURNER199067}.

Although the Sun is a poor source of $^{57}$Fe axions due to its cool core $T_8 \approx 0.15$, the suppressed Boltzmann occupation of the excited state is mitigated by the Sun's proximity. The thermally broadened line of 
$^{57}$Fe axions can then be detected on Earth \cite{PhysRevLett.75.3222} by the inverse resonant process $a+{^{57}\mathrm{Fe}} \rightarrow {^{57}\mathrm{Fe}^*} \rightarrow {^{57}\mathrm{Fe}} + \gamma$. Searches have been conducted by the CAST Collaboration \cite{CASTcollaboration_2009} as well as other groups \cite{KRCMAR199838,NAMBA2007398,Derbin,CUORE:2012ymr,DiLuzio:2021qct}.
The ``axion helioscope" detector exploits $a \rightarrow \gamma$ conversion as axions pass through a strong magnetic field. White dwarfs possessing high core temperatures and strong magnetic fields have also been used as sources and converters of $^{57}$Fe axions \cite{Fleury:2022plh}.

Haxton and Lee identified another isotope, $^{23}$Na, as the dominant axion pump for high temperatures $T_8 \approx 10$, though this observation has drawn little attention.  Here we show that $^{23}$Na is the strongest continuous source of line axions in the galaxy.  As these axions traverse the galactic magnetic field, they produce a distinctive gamma ray signal, ideal for telescopes like COSI \cite{Tomsick:2023aue}.

The ALP-nucleon interaction is
\begin{eqnarray}
\mathcal{L} &=& {1 \over 2m_N}  \bar{N}\gamma^\mu \gamma_5 (g^0_{aNN} +g^3_{aNN} \tau_3 ) N \, \partial_\mu a  \\
	&=& {1 \over 2m_N} \bar{N}\gamma^\mu \gamma_5 \left[ g_{app}\textstyle{ \left({1+\tau_3 \over 2} \right) }+ g_{ann} \left({1-\tau_3 \over 2} \right) \right] N \, \partial_\mu a \nonumber
\label{eq:ncoupling}
\end{eqnarray}
where $\tau_3$ is the third component of isospin, with $g_{app}= g^0_{aNN}+g^3_{aNN}$ and $g_{ann}=g^0_{aNN}-g^3_{aNN}$.  
For the KSVZ \cite{PhysRevLett.43.103,SHIFMAN1980493} and DFSZ \cite{DINE1981199,osti_7063072} QCD axions,
the couplings are related to the axion mass scale $f_a$ \cite{Cortona},
\begin{eqnarray}
	\left. \begin{array}{l}  g_{app}^\mathrm{KSVZ}  \\[3pt] g_{ann}^\mathrm{KSVZ}  \end{array} \right\} &=& {m_N \over f_a}  \left\{ \begin{array}{r}  -0.47  \\ -0.02 \end{array}  \right.   \nonumber \\
	\left. \begin{array}{l}  g_{app}^\mathrm{DFSZ}  \\[3pt] g_{ann}^\mathrm{DFSZ}  \end{array} \right\} & = & {m_N \over f_a}  \left\{ \begin{array}{r}  -0.182-0.435 \sin^2{\beta}  \\ -0.160+0.414 \sin^2{\beta} \end{array}  \right.   
\end{eqnarray}
where the angle $\beta$ is defined as in \cite{Caputo:2024oqc}. 
The constraint derived from axion cooling of SN1987A \cite{Caputo:2024oqc}
\begin{equation}
	\left[ (g_{app}+0.435g_{ann})^2+1.45g_{ann}^2 \right]^{1/2} < 1.19 \times 10^{-9}
\end{equation}
sets the scale of $10^{-9}$ used in this study.

\begin{table}[h!]
	\centering
	\begin{tabular}{ccccc}
		\hline
		& & & & \\[-4pt]
		Source~ &~$\Delta$E (keV) & T range &~ B(M1) & ~$g_{aNN}^\mathrm{eff}$~ \\[4pt]
		\hline
		& & & & \\[-4pt]
		$^{57}$Fe & 14.4 & $\lesssim 1.5 T_8$ & 0.0082 & $g_{ann}+0.088\,g_{app}$ \\
		$^{23}$Na & 440.2 & $\gtrsim 4.1 T_8$ & 0.225 & $g_{app}-0.062\,g_{ann}$ \\[4pt]
		\hline
	\end{tabular}
	\caption{Comparative properties of the axion emitters $^{57}$Fe and $^{23}$Na. The temperature range defines where these
		metals could dominate axion emission from stars.}
	\label{tab:one}
\end{table}

Equation (\ref{eq:ncoupling}) generates a nuclear amplitude proportional to the Pauli spin operator, so that the allowed nuclear transitions are M1. To minimize nuclear structure uncertainties,  the axion decay rate is expressed in terms of the known gamma decay rate  \cite{PhysRevLett.66.2557},
\begin{equation}
	{\omega_a \over \omega_\gamma} = {1 \over 2 \pi \alpha} {1 \over 1+ \delta^2} \left[ {g^0_{aNN} \beta+ g^3_{aNN} \over (\mu_0-{1 \over 2}) \beta +\mu_1 -\eta} \right]^2
	\label{eq:rate}
\end{equation}
where $\delta$ is the E2/M1 mixing ratio, and $\mu_0=0.88$ and $\mu_1=4.706$ are the isoscalar and isovector nucleon magnetic moments, respectively.  The nuclear physics input is limited to two ratios of reduced matrix elements
\begin{equation}
\beta \equiv  {
\langle J_f \|  \sum\limits_{i=1}^A \boldsymbol\sigma(i)  \| J_i \rangle
\over 
\langle J_f \| \sum\limits_{i=1}^A \boldsymbol\sigma(i) \,\tau_3(i) \| J_i \rangle 
},
~~
\eta \equiv - {
\langle J_f \| \sum\limits_{i=1}^A \boldsymbol\ell(i)\, \tau_3(i) \| J_i \rangle 
\over 
\langle J_f \| \sum\limits_{i=1}^A \boldsymbol\sigma(i)\, \tau_3(i) \| J_i \rangle 
}
\end{equation}
which were evaluated in \cite{PhysRevLett.66.2557} for cases of interest.

This mechanism requires a low-lying excited state that can be thermally populated under stellar conditions.
For example, the Boltzmann occupation of the $^{57}$Fe 14.4 keV state at red giant temperatures ($T_8 \approx 1$) is 0.27. 
The corresponding occupation at solar temperatures, obtained by averaging over the solar core, is just $5 \times 10^{-6}$.

Among the three axion emitters discussed in \cite{PhysRevLett.66.2557}, $^{23}$Na has the most favorable nuclear physics.  Table \ref{tab:one} compares $^{23}$Na and $^{57}$Fe. The $^{23}$Na  B(M1) value -- the magnetic transition probability -- is a significant fraction of a single-particle unit, nearly 30 times stronger than in $^{57}$Fe.  In addition, $^{23}$Na has an unpaired proton, while $^{57}$Fe has a valence neutron.    Proton couplings are favored: the KSVZ axion gives  $(g_{app}/g_{ann})^2 \approx 550$, while the ratio can range up to infinity for the DFSZ axion, depending on ${\beta}$.  The shell-model prediction for the effective nucleon coupling in $^{23}$Na is
$ g_{aNN}^\mathrm{eff}  = g_{app}-0.062 \,g_{ann}$.

$^{23}$Na also has significant astrophysical advantages.   First is its abundance.  The Sun's $^{57}$Fe is primordial, incorporated at the time of formation: the core contains $7.7 \times 10^{-6} M_\odot$ of $^{57}$Fe.  By contrast, $^{23}$Na is produced {\it in situ} in carbon-burning massive stars, at temperatures
$7 \lesssim T_8 \lesssim 13$.  The total mass synthesized is typically $\approx$ 0.1 $M_\odot$.  As these temperatures are maintained for times that range up to 60,000 y, several hundred galactic sources may be producing axions today.  Second, unlike solar $^{57}$Fe axions, galactic $^{23}$Na axions traverse the magnetic fields of the progenitor star and the galaxy, allowing some of the axions to convert to detectable gamma rays. 

It is very rare for stars to synthesize odd-A isotopes in large quantities -- the nuclear astrophysics favors tightly bound $\alpha$-stable nuclei like C, O, and Ne -- and still rarer for such isotopes, once produced, to evade rapid burn-up in $T_8 \approx 10$ stellar plasmas.  $^{23}$Na is an exception.  The isotope is produced during carbon burning,
\begin{equation}
	^{12}\mathrm{C} + {^{12}\mathrm{C}} \rightarrow {^{24}\mathrm{Mg}}^* \rightarrow \left\{ \begin{array}{l} ^{20}\mathrm{Ne} +{^{4}\mathrm{He} }+ 4.62 \mathrm{~MeV} \\
		^{23}\mathrm{Na} + p  ~\,~~ +2.24 \mathrm{~MeV} \\ ^{23}\mathrm{Mg} + n~\,~ -2.60 \mathrm{~MeV} \end{array} \right.
\end{equation}
in the late-stage evolution of massive stars.  
The stars that prove to be of most interest are in the mass range of $7.5-15$ $M_\odot$,  the progenitors of ONeMg white dwarfs (WDs) and of electron capture (EC)
and core-collapse (CC) supernovae (SNe) \cite{Limongi_2024}.

The stellar input for the subsequent galactic modeling of $^{23}$Na axion production was generated using the MESA code \cite{Paxton2011,Paxton2013,Paxton2015,Paxton2018,Paxton2019,Jermyn2023}.
SN progenitors of mass 9, 10, ..., 30 were evolved through their various burning stages, including C, Ne, and O burning. The output
includes the photon, neutrino (including Urca production from $^{23}$F/$^{23}$Ne/$^{23}$Na and $^{25}$Ne/$^{25}$Na/$^{25}$Mg), and axion luminosities, as well as the radial profiles of composition and temperature at each time slice. The axion emission is computed as a function of time and radial coordinate
from the $^{23}$Na mass fraction and Boltzmann occupation of the  $5/2^+$ 440 keV level of $^{23}$Na.  The axion
emission rate is computed from Eq. (\ref{eq:rate}), with $\omega_\gamma=6.10 \times 10^{11}$/sec and $\delta=0.065$ taken from ENSDF \cite{SHAMSUZZOHABASUNIA20211}.

\begin{figure}
\centering
{
\includegraphics[width=\linewidth]{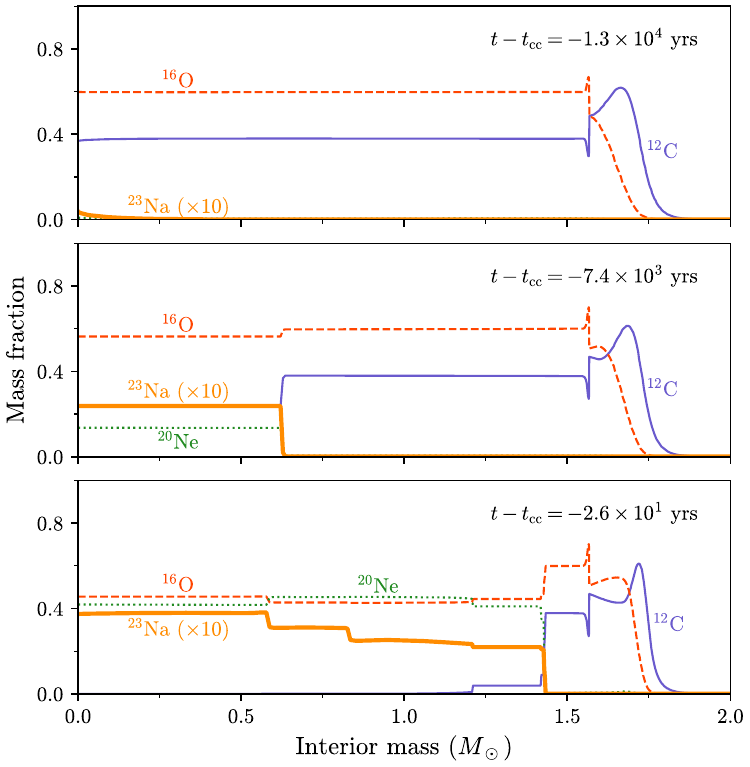}
\caption{Mass fractions of C, O, Ne, and $^{23}$Na ($\times10$) at the start of C+C burning (top), at the end of C burning in the innermost zone, and at the end of C burning (bottom) for a 9$M_\odot$ SN progenitor.}
\label{fig:one}
}
\end{figure}

The amount of $^{23}$Na synthesized in the inner cores of these stars is large, 
ranging from 0.044$M_\odot$ in a $9M_\odot$ star
to 0.13$M_\odot$ in a 30$M_\odot$ one.  The distributions of $^{23}$Na before and after C burning are shown in Fig.~\ref{fig:one}.
The core temperatures at the time $^{23}$Na reaches its peak abundances range from $T_8 \approx 12$ in a $9M_\odot$ progenitor to $T_8 \approx 24$ in a $30M_\odot$ star.  
Most importantly, C burning times are much longer in lighter progenitors: in a $9M_\odot$ SN star, burning begins when the core temperature reaches $T_8 \approx 7$ and continues for $\approx$ 12,500 y.   By contrast, C burning lasts about 25 y in a $30M_\odot$ progenitor.  Lower mass stars dominate 440 keV axion production because of their long C-burning times and greater abundance.  
The galactic 440 keV axion flux is continuous, as source durations are long compared to SN and ONeMg WD birth rates. The Boltzmann factors are favorable: at the end of $^{23}$Na synthesis in a $9M_\odot$ star, the 440 keV state occupation probability is 0.11.  

To estimate the flux at Earth we constructed a galactic model with which we can perform Monte Carlo sampling on the position, mass, and age of contributing SN progenitors. To describe the progenitor position in the Milky Way, we use a version of the Stecker and Jones distribution \cite{1977ApJ...217..843S}, as parameterized in \cite{Verberne:2021tse}, to match the current inventory of SN remnants.  These remnants provide a partial record of Milky Way SN activity over the last $\sim$ 100,000 years. We altered this distribution to produce a nonzero remnant density in the bulge, bringing the distribution into better agreement with that of \cite{Case_1998} near the galactic center.  
The distribution depends only on $r$ (in kpc), the distance from the galactic center,  
\begin{equation}
\rho(r) = N \left( {r + 0.638 \over r_\odot} \right)^{0.74}  \exp{\left(  -3.54 {(r-r_\odot) \over r_\odot} \right)} 
\end{equation}
where $r_\odot \approx 8.5$ kpc.  The constant $N$ is fixed by the requirement that $\rho(r)$ be a probability density.

Because we treat a flux that falls off as $1/r^2$, the thickness of the disk must be included to keep integrations well behaved.  We describe the $z$-dependence by
\begin{equation}
P(z) = {1 \over 2h} \exp{\left(-{|z| \over h} \right) }
\end{equation}
setting the scale height $h=0.4$ kpc while placing the Earth at the midpoint of the disk.  While our approach is reasonable given other uncertainties, improvements could be made by adapting more sophisticated galactic models that account for the nonuniform distribution of stars \cite{KM}.

Progenitors with masses $\gtrsim 9.2 M_\odot$ produce standard core-collapse (CC) SNe \cite{Limongi_2024}. After completion of C burning, stars just below this cutoff develop an inert ONeMg electron-degenerate core, with several possible outcomes. Those stars that retain most of their envelopes experience electron capture (EC) on Ne, and can explode either as a thermonuclear ECSN or a CCSN  \cite{Limongi_2024}.  We group
these stars together as the SN population, using $8.5 M_\odot$ as the lower bound on the progenitor mass (see below).

In the sampling over progenitor mass, we use an initial mass function (IMF) of the Salpeter form \cite{1955ApJ...121..161S}, $P(M) \propto M^{-2.35}$.  To each progenitor of mass $M$ we assign a probability given by the integral of the IMF over $M \pm 0.5M_\odot$.  The Salpeter IMF provides a good fit to the contemporary inventory
of SN remnants \cite{Williams_2018}.

The sampling over time requires us to adopt a rate for SNe.  Various estimates of the CCSN rate for the galaxy as a whole have been made in recent years, with typical results being 1.63$\pm$0.46/century (aggregated data) \cite{ROZWADOWSKA2021101498}, 1.95 $\pm$ 0.41/century (extragalactic survey of similar galaxies) \cite{Planck}, and 1.9 $\pm$ 1.1/century ($^{26}$Al) \cite{Diehl}.  Use of these rates is complicated by the fact that, due to the $1/r^2$ falloff of the axion flux, 50\% axions detected on Earth will originate from within 2.1 kpc.  The rate  may be higher in the spiral arms, where our solar system resides \cite{10.1046/j.1365-8711.1999.02145.x}.   We adopt a rate of CC and EC SNe  of 2.5/century: the results below can be scaled to any other value.  The time over which we integrate is set by the duration of C burning in the 9$M_\odot$ progenitor, 12,500y.   We create an ensemble of $\approx$ 300 progenitors that are within 12,500y of exploding,  with Poisson statistics determining the actual number and distribution in time. The mass and location of each progenitor is determined by Monte Carlo.  The axion flux at Earth for the ensemble is then computed.  By repeating the calculation for 1000 ensembles, we determine the flux probability distribution.

A second important axion source is ONeMg white dwarfs (WDs), with masses in the range $\approx$ 1.1--1.4 $M_\odot$. In the fine-grid study of  \cite{Limongi_2024}, progenitors with masses of 7.5--9.20 $M_\odot$ were found to produce degenerate ONeMg cores, with those with  $M \ge 8.5 M_\odot$ ending their evolution as ECSNe and those with $M \approx 7.5-8.5 M_\odot$ becoming ONeMg WDs.  Certain heavier progenitors can also evolve to ONeMg WDs depending on the description of mass loss.  Because they evolve to similar final states, we represent all WD progenitors by a single star, a 11$M_\odot$ MESA progenitor that produces a 1.134$M_\odot$  ONeMg WD containing $\approx$ 0.07$M_\odot$ of ${^{23}\mathrm{Na}}$.  An improved approach would be to generate a finely spaced grid of progenitors, a task beyond our scope, but might be within that of the modeling program described in  \cite{Limongi_2024}.

An ONeMg WD continues axion production for $\approx 10^4$y after C burning \cite{Camisassa}, as gravitational energy released through contraction keeps core temperatures high despite neutrino losses.   This further extends the axion production, which can be five times that of comparable light SN progenitors. In our MESA calculation, axion production extends over 40,000y (15,000y) in the C burning (cooling) phase, accounting for 84\% (15\%)  of the axion fluence.

Given a ONeMg WD birth rate, one can calculate the galactic production, just as was done for SNe.  For the ONeMg WD progenitor mass range of  \cite{Limongi_2024}, 7.5--8.5$M_\odot$,
the IMF and SN rate can be combined to yield a birth rate of 1/175y. 
Alternatively, the star formation rate of $\approx$ 1/1.9y and estimated heavy (1.1--1.4$M_\odot$) WD fraction (1--2\%)
\cite{Cunningham} together yield a birth rate of 1/(100--200)y. In our simulations we use 1/175y as a representative rate.

\begin{figure}
	\centering
	\includegraphics[scale=0.205]{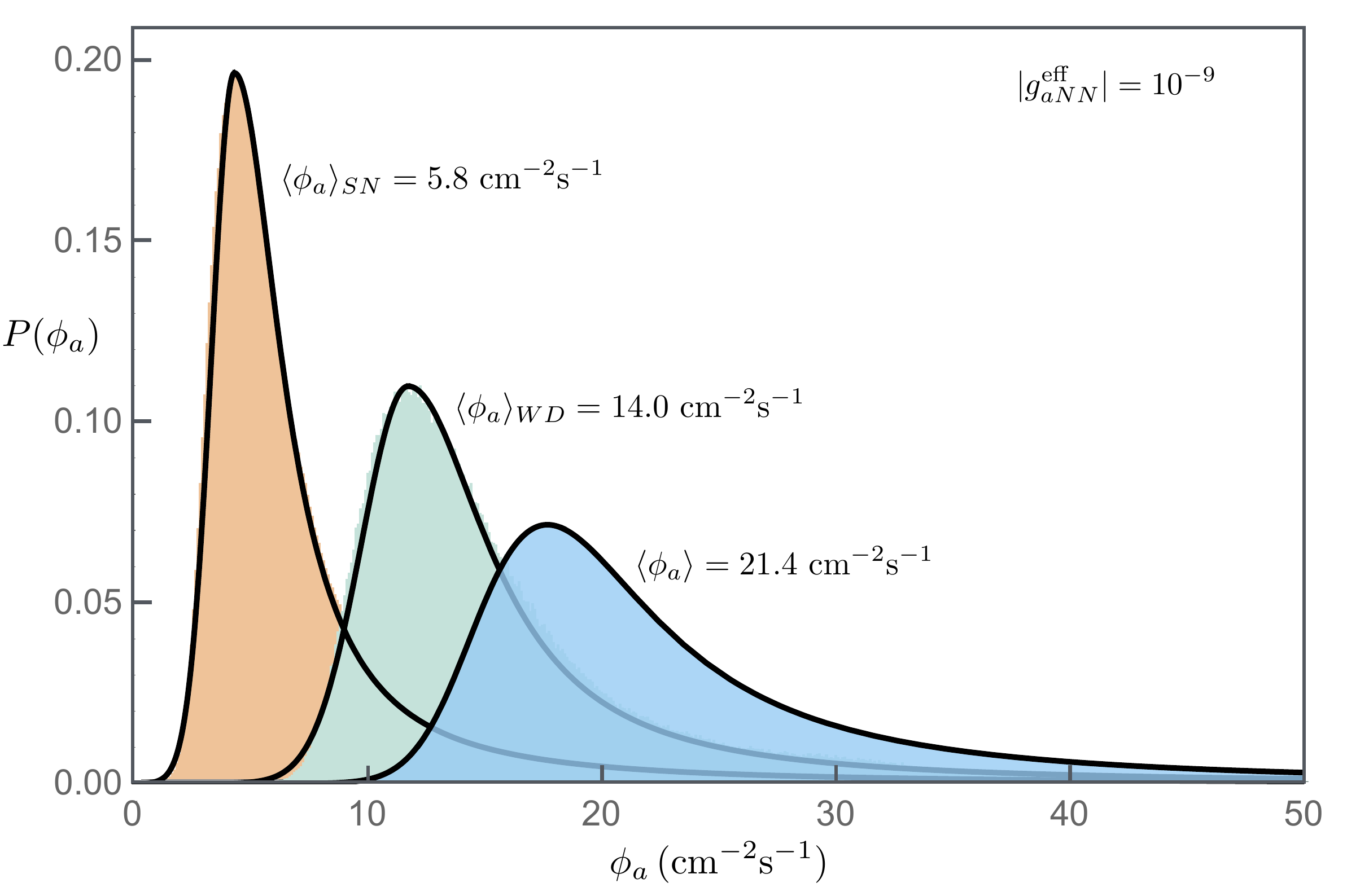}
	\caption{From left to right, the Monte Carlo axion flux probability distributions for SN and WD sources separately, and combined.  The solid lines are the analytic fits (see text).}
	\label{fig:two}
\end{figure}

The resulting SN and WD flux probabilities 
$P(\phi_a)$ computed for $g_{a NN}^\mathrm{eff} = 10^{-9}$ are shown in Fig.~\ref{fig:two}.  As several hundred progenitors contribute, the distributions have defined lower boundaries.  However, the
high-flux tails are extended, reflecting the random nature of near-Earth sources.  The Monte Carlo results can be approximated by a 
two-sided Gaussian-Lorentzian
\begin{equation}
	P(\phi_a) = \sqrt{\pi \over 2} {1 \over \sigma_1 + \sigma_2} \left\{ \begin{array}{ll} e^{-(\phi_a-\phi_0)^2 /2 \sigma_1^2} & \phi_a < \phi_0 \\
		\left[ 1 + \pi {(\phi_a-\phi_0)^2 \over 2 \sigma_2^2} \right]^{-1} & \phi_a>\phi_0 \end{array} \right.
\end{equation}
where $\phi_0$ is the most probable flux.  We find $\{\phi_0,\sigma_1,\sigma_2\} = \{11.8,1.99,5.32\}, \{4.28,0.98,3.08\}$, and $ \{17.3,2.63,8.52\}$ cm$^{-2}$s$^{-1}$ for the SN, WD, and total fluxes, respectively.  Because the probability profile is skewed, we will use the median
to define a representative
flux for the exploratory study below, $\langle \phi_a \rangle= 21.4$ cm$^{-2}$s$^{-1}$.

Despite the lower birthrate of ONeMg WDs, they account for $\approx$ 70\% of the flux because of their long duration.
$\sigma_1$ and $\sigma_2$ determine the probabilities of measuring a flux above or below the most probable value $\phi_0$, respectively.  As $\sigma_2 \gg \sigma_1$, a measurement is more likely to find a value in the Lorentzian tail than otherwise.\\
~\\
{\it ALP conversion in the galactic magnetic field:} Naively one expects a galactic flux of axions to be less important than
a solar flux by $(d_\mathrm{sun}/d_\mathrm{gal})^2 \approx 10^{-19}$.  In fact, the factors previously discussed yield a galactic $^{23}$Na to solar
$^{57}$Fe flux ratio at Earth of 1/350 (KSVZ couplings).  When in addition one considers the conversion of light ALPS to detectable 
$\gamma$s, the factor
$(B_T d)^2$ swings the balance sharply in favor of the galaxy,  which enjoys an advantage of  $\approx 10^{17}$ over the 9.26\,m, 9\,T CAST magnet.

$^{23}$Na axions arrive on Earth after transiting
both the local magnetic field of the progenitor and that of the galaxy.  Magnetic fields can convert the axions to 440 keV $\gamma$s 
through the coupling
\begin{equation} 
	\mathcal{L} =-g_{a \gamma \gamma} \vec{E} \cdot \vec{B} \, a 
\end{equation}
The CAST limit on $g_{a \gamma \gamma}$ is \cite{CAST}
\begin{equation}
{~|g_{a \gamma \gamma}|  < 0.66 \times 10^{-10}\,\,(95\%\, \rm{C.L.}}) ~\mathrm{for}~m_a \lesssim 0.02 \mathrm{\, eV}.
\end{equation}
Constraints from axion energy loss in horizontal branch
and AGB star produce similar bounds \cite{Caputo:2024oqc,Carenza:2024ehj}. Tighter limits have been established by ADMX
\cite{PhysRevLett.120.151301,PhysRevLett.124.101303,PhysRevLett.127.261803} and from analyses of the TeV $\gamma$-ray source Markarian 421 \cite{LI2024139075}, but they are valid only in specific mass ranges.

As axion couplings to quarks and gluons also imply axion-pion mixing, one expects $|g_{a\gamma \gamma}| \sim ({\alpha/ m_N}) |g_{aNN}|$.  For the KSVZ axion, this relates a coupling of the size used here $|g_{app}| \sim 10^{-9}$ to $|g_{a\gamma \gamma}| \sim 4.9 \times 10^{-12}$ GeV$^{-1}$. Consequently, in the work below, we take the nominal value of $g_{a\gamma \gamma}$ to be $5 \times 10^{-12}$, placing both
couplings near their current experimental limits.  Observational 
signals are governed by $|g^\mathrm{eff}_{aNN} g_{a \gamma \gamma}|$, which then has the nominal value $5 \times 10^{-21}$.  We will explore below the potential of COSI to probe couplings below this nominal value.

The conversion probability of very light $^{23}$Na ALPs to $\gamma$s depends on the component of the galactic magnetic field
perpendicular to the ALP's direction.  Both the in-plane transverse component and that perpendicular to the Milky Way disk contribute. Here we simplify, taking the total transverse component to be constant at $B_T \approx 1\, \mu G$.  This choice is consistent with retaining just the disk-perpendicular component of the ``logistic" parameterization from Fig.~16 of \cite{Unger_2024}.

For an axion of energy $E$ traveling a distance $d$ through a homogeneous field, the conversion probability is~\cite{PhysRevD.37.1237}
\begin{equation}
	P_{a \to \gamma} = (\Delta_{a \gamma} d)^2 \,\frac{\sin^2(\Delta_{\rm osc}d/2)}{(\Delta_{\rm osc}d/2)^2}
\label{eq:prob}
\end{equation} 
where $\Delta_{\rm osc}$ = $\left[(\Delta_a - \Delta_{\rm pl})^2 +4\Delta^2_{a \gamma}\right]^{1/2}$, $\Delta_{a \gamma}=g_{a \gamma \gamma} B_T/2$, $\Delta_a = -m^2_a/2E$, and $\Delta_{\rm pl} = -\omega^2_{\rm pl}/2E$.
Using an galactic electron density $n_e \approx 1.1 \times 10^{-2}$ cm$^{-3}$, we obtain
the plasma frequency  $\omega_{\rm pl} = \sqrt{4 \pi \alpha n_e/m_e} \approx 3.9 \times 10^{-12}$ eV.
The various terms can be expressed parametrically as
\begin{eqnarray}
	\Delta_a &=& -7.8 \times 10^{-3} \times \left(\frac{m_a}{10^{-11}\,\rm {eV}}\right)^2 \left(\frac{E}{\rm MeV}\right)^{-1} \rm kpc^{-1} \nonumber\\
	\Delta_{\rm pl} &=& -1.1 \times 10^{-3} \times \left(\frac{E}{\rm MeV}\right)^{-1} \rm kpc^{-1} \\
	\Delta_{a \gamma} &=& 1.5 \times 10^{-2} \times\left(\frac{g_{a \gamma \gamma}}{10^{-11}\,\rm GeV^{-1}}\right) \left(\frac{B_T}{10^{-6}\,\rm G}\right) \rm kpc^{-1} \nonumber
\end{eqnarray}

The $\gamma$-ray flux is obtained by repeating the Monte Carlo galaxy simulation with the conversion probability included. Unlike the axion flux, where nearby sources are especially important, the additional factor of $d^2$ from Eq. (\ref{eq:prob}) leads to a source weighting proportional to the number of sources at a distance $d$, favoring distant sources.  The contribution within a given solid angle is then proportional to the number of stars in the associated cone.

\begin{figure}
\centering
\includegraphics[scale=0.205]{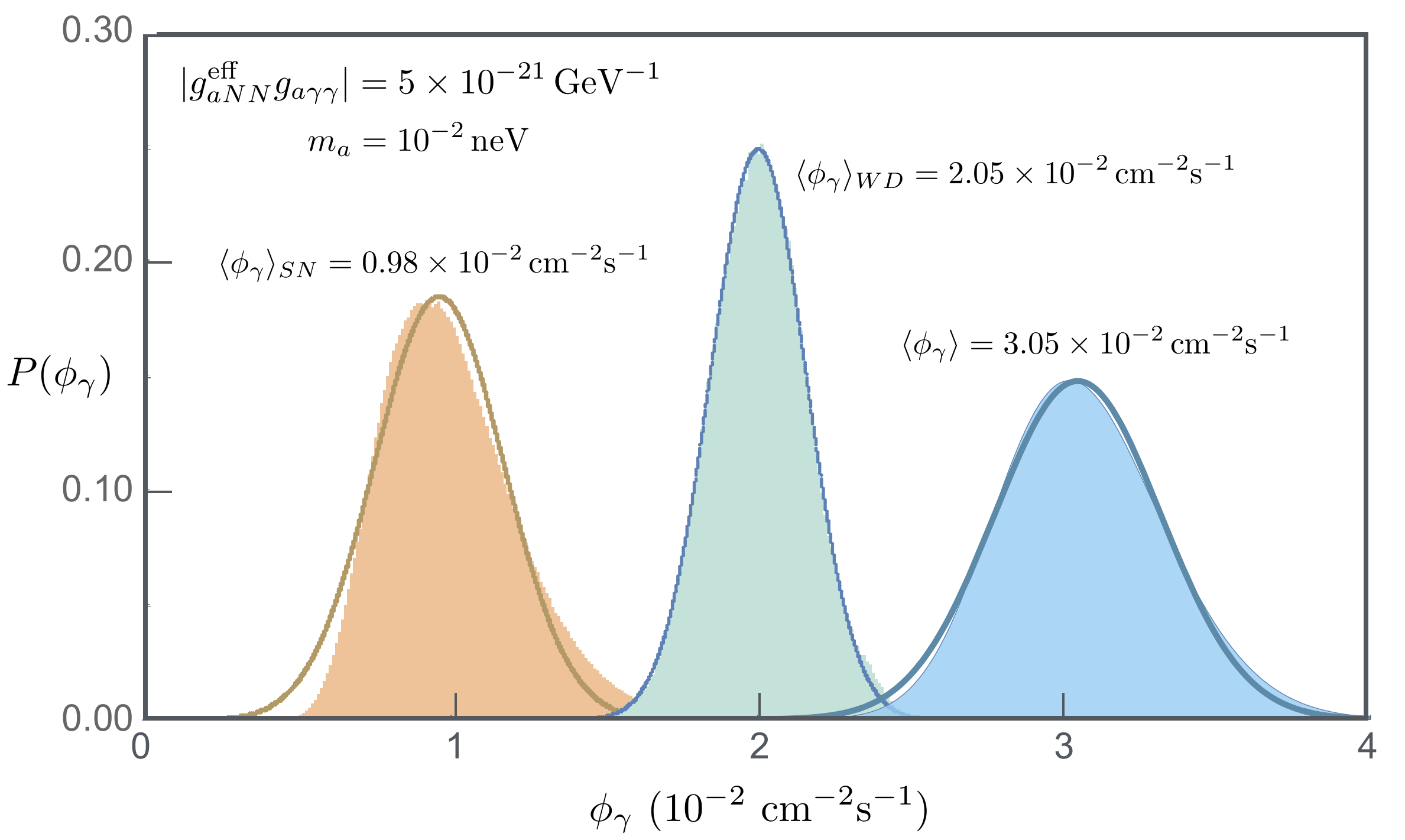}
\caption{As in Fig 2, but for the conversion $\gamma$s.  The Monte Carlo
results are shown with simple Gaussian fits (solid lines).}
\label{fig:three}
\end{figure}

The $\gamma$-ray flux probability distributions are shown in Fig.~\ref{fig:three} for a light ALP ($m_a = 10^{-11}$ eV). We find $\langle \phi_\gamma \rangle \approx 3.05 \times 10^{-2}$ cm$^{-2}$s$^{-1}$ and a conversion probability $\langle P_{a \rightarrow \gamma} \rangle\approx  0.0013$. $P(\phi_\gamma)$ is well described by a Gaussian (see Fig.~\ref{fig:three}) with $\sigma \approx 0.27 \times 10^{-2}$ cm$^{-2}$s$^{-1}$.  As the conversion occurs far from the axion source, the $\gamma$ line profile should be narrow, with a Gaussian width of $\approx$ 0.9 keV, reflecting the thermal line broadening of their axion parents.\\

\begin{figure}
\centering
\includegraphics[scale=0.43]{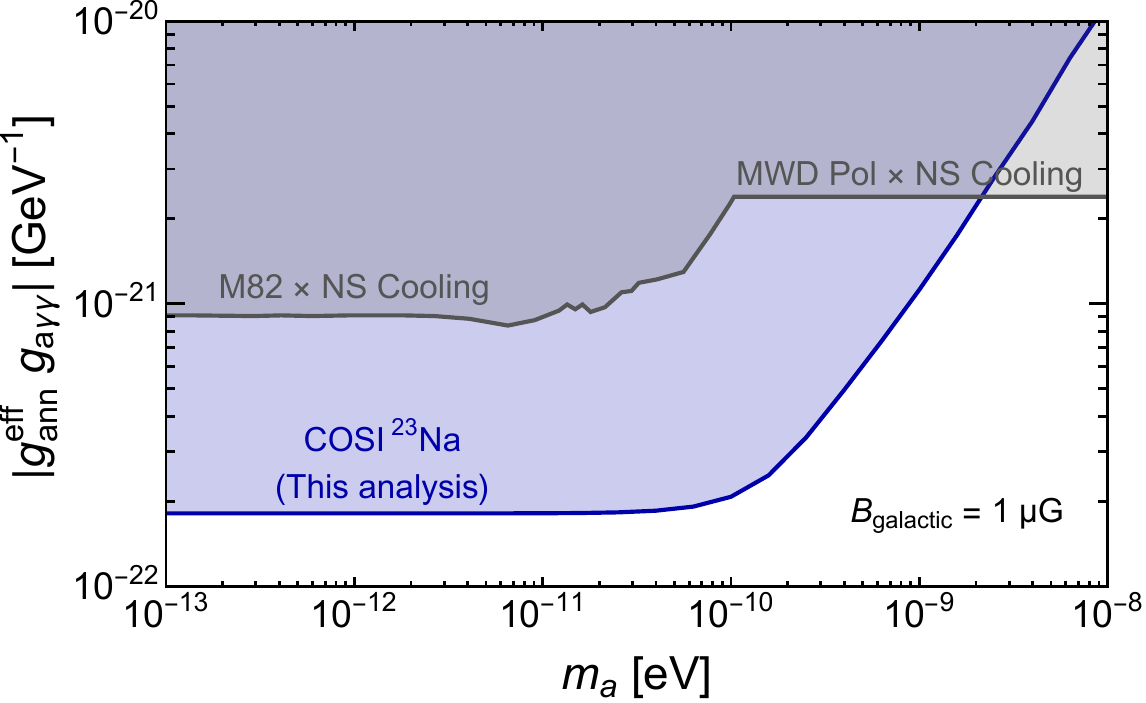}
\caption{Expected $3\sigma$ limit on the ALP coupling $|g_{aNN}^\mathrm{eff} g_{a \gamma \gamma}|$ from a COSI search
for $^{23}$Na $a \rightarrow \gamma$ conversion, after two years of surveying (blue shaded). The other constraints
shown (gray shaded) were obtained by combining leading limits on $g_{a \gamma \gamma}$, from axion-photon conversion in M82 and axion-induced polarization measurements in magnetic white dwarfs~\cite{Ning:2024eky,Benabou:2025jcv}, with the bound on $g_{aNN}^\mathrm{eff}$ obtained from isolated neutron star cooling~\cite{Buschmann:2021juv}.}
\label{fig:four}
\end{figure}

\noindent
{\it $\gamma$-ray detection with the COSI detector:}  COSI's planned launch is 2027 \cite{Tomsick:2023aue}. As an all-sky detector, it is optimized for continuous $\gamma$-ray sources at $\approx$ 1 MeV. From Fig.~2(a) of \cite{Tomsick:2023aue} we obtain a COSI narrow-line point source sensitivity of $6 \times 10^{-6}\,\mathrm{cm}^{-2}\,\mathrm{s}^{-1} \, (3 \sigma)$ at 440 keV after two years of survey time. This sensitivity can be adapted to the case of multiple point sources by accounting for the background noise scaling of $\sqrt{\Omega/\Omega_0}$, where $\Omega_0$ is the solid angle subtended by a $4^{\circ}$ cone, COSI's resolution.  This cone spans the effective height of the galactic plane, but not the distribution within the plane: as 90\% of the signal comes from the hemisphere toward the galactic center, we estimate that $\sqrt{\Omega/\Omega_0} \approx 6.7$, yielding an effective COSI sensitivity to $^{23}$Na conversion $\gamma$s of $4 \times 10^{-5} \mathrm{cm}^{-2}\mathrm{s}^{-1} \, (3 \sigma)$.  Our naive analysis makes no use of the potentially helpful correlation of signal with angular stellar density.

Figure~\ref{fig:four} shows the resulting bounds on the ALP coupling $|g_{aNN}^\mathrm{eff} g_{a \gamma \gamma}|$ as a function
of the ALP mass.  A COSI sensitivity
of $4 \times 10^{-5} \mathrm{cm}^{-2}\mathrm{s}^{-1}$ will probe $|g_{aNN}^\mathrm{eff} g_{a \gamma \gamma}|$ to $1.8 \times 10^{-22}$ GeV$^{-1}$ for sufficiently light ALP masses
($m_a \lesssim 0.1$ neV). This limit corresponds to a $3\sigma$ detection at the level defined by $\langle \phi_\gamma \rangle$.  
The figure also shows the $|g_{aNN}^\mathrm{eff} g_{a \gamma \gamma}|$ constraints that can be obtained by combining the leading astrophysical bounds on $g_{a \gamma \gamma}$, from axion-photon conversion in M82 and axion-induced polarization measurements in magnetic white dwarfs~\cite{Ning:2024eky,Benabou:2025jcv}, with the bound on $g_{aNN}^\mathrm{eff} = g_{app} -0.062 g_{ann}$ obtained from the cooling of isolated neutron stars~\cite{Buschmann:2021juv}.
\\
~\\
{\it Outlook:}  The conversion $\gamma$s from galactic $^{23}$Na axions appear to be an interesting target for COSI and similar all-sky detectors.  There are further issues to be explored.  First, the exploratory calculations that we have performed could be improved by utilizing more sophisticated models of the star \cite{KM} and magnetic field \cite{Unger_2024} distributions. Second, axion interactions with the progenitor's magnetic field can enhance the $a \rightarrow \gamma$ conversion, extending sensitivities to heavier ALPs \cite{PhysRevLett.133.211002}. As much is known about
the distribution of field strengths in WDs, this physics could be incorporated into the Monte Carlo study. Third, axion
emission during C burning can affect progenitor evolution. While our preliminary studies suggest that the effects are important only if $|g_{a N N}^\mathrm{eff}| \gtrsim 10^{-8}$, still this should be quantified.

\begin{acknowledgments}
We thank Steve Boggs, Liang Dai, Chris Fryer, Georg Raffelt, Josiah Schwab, and Stan Woosley for helpful discussions.  
We thank the MESA Collaboration for their open source software; our modifications are documented on GitHub \cite{GitHub}. 
WH and AR are supported through the NSF cooperative agreement 2020275.   WH is also supported by the US DOE (DE-SC0004658 and DE-SC0023663) and by the Heising-Simons Foundation (00F1C7).  The progenitor modeling was performed by AM (2023) and XL (2024) as undergraduate research projects for the NSF Center N3AS.
\end{acknowledgments}

\bibliographystyle{apsrev4-1}
\bibliography{Na23.bib}

@article{PhysRevLett.38.1440,
  title = {$\mathrm{CP}$ Conservation in the Presence of Pseudoparticles},
  author = {Peccei, R. D. and Quinn, Helen R.},
  journal = {Phys. Rev. Lett.},
  volume = {38},
  issue = {25},
  pages = {1440--1443},
  numpages = {0},
  year = {1977},
  month = {Jun},
  publisher = {American Physical Society},
  doi = {10.1103/PhysRevLett.38.1440},
  url = {https://link.aps.org/doi/10.1103/PhysRevLett.38.1440}
}

@article{TURNER199067,
title = {Windows on the axion},
journal = {Physics Reports},
volume = {197},
number = {2},
pages = {67-97},
year = {1990},
issn = {0370-1573},
doi = {https://doi.org/10.1016/0370-1573(90)90172-X},
url = {https://www.sciencedirect.com/science/article/pii/037015739090172X},
author = {Michael S. Turner}
}

@article{PhysRevLett.40.223,
  title = {A New Light Boson?},
  author = {Weinberg, Steven},
  journal = {Phys. Rev. Lett.},
  volume = {40},
  issue = {4},
  pages = {223--226},
  numpages = {0},
  year = {1978},
  month = {Jan},
  publisher = {American Physical Society},
  doi = {10.1103/PhysRevLett.40.223},
  url = {https://link.aps.org/doi/10.1103/PhysRevLett.40.223}
}

@article{PhysRevLett.40.279,
  title = {Problem of Strong $P$ and $T$ Invariance in the Presence of Instantons},
  author = {Wilczek, F.},
  journal = {Phys. Rev. Lett.},
  volume = {40},
  issue = {5},
  pages = {279--282},
  numpages = {0},
  year = {1978},
  month = {Jan},
  publisher = {American Physical Society},
  doi = {10.1103/PhysRevLett.40.279},
  url = {https://link.aps.org/doi/10.1103/PhysRevLett.40.279}
}

@article{Caputo:2024oqc,
    author = "Caputo, Andrea and Raffelt, Georg",
    title = "{Astrophysical Axion Bounds: The 2024 Edition}",
    eprint = "2401.13728",
    archivePrefix = "arXiv",
    primaryClass = "hep-ph",
    reportNumber = "MPP-2024-13, CERN-TH-2024-013",
    doi = "10.22323/1.454.0041",
    journal = "PoS",
    volume = "COSMICWISPers",
    pages = "041",
    year = "2024"
}

@article{PhysRevLett.128.091102,
  title = {Upper Limit on the QCD Axion Mass from Isolated Neutron Star Cooling},
  author = {Buschmann, Malte and Dessert, Christopher and Foster, Joshua W. and Long, Andrew J. and Safdi, Benjamin R.},
  journal = {Phys. Rev. Lett.},
  volume = {128},
  issue = {9},
  pages = {091102},
  numpages = {9},
  year = {2022},
  month = {Mar},
  publisher = {American Physical Society},
  doi = {10.1103/PhysRevLett.128.091102},
  url = {https://link.aps.org/doi/10.1103/PhysRevLett.128.091102}
}

@article{Raffelt:2006cw,
	author = "Raffelt, Georg G.",
	editor = "Kuster, Markus and Raffelt, Georg and Beltran, Berta",
	title = "{Astrophysical axion bounds}",
	eprint = "hep-ph/0611350",
	archivePrefix = "arXiv",
	reportNumber = "MPP-2006-172",
	doi = "10.1007/978-3-540-73518-2_3",
	journal = "Lect. Notes Phys.",
	volume = "741",
	pages = "51--71",
	year = "2008"
}

@article{PhysRevD.94.085012,
  title = {Probing axions with the neutrino signal from the next Galactic supernova},
  author = {Fischer, Tobias and Chakraborty, Sovan and Giannotti, Maurizio and Mirizzi, Alessandro and Payez, Alexandre and Ringwald, Andreas},
  journal = {Phys. Rev. D},
  volume = {94},
  issue = {8},
  pages = {085012},
  numpages = {19},
  year = {2016},
  month = {Oct},
  publisher = {American Physical Society},
  doi = {10.1103/PhysRevD.94.085012},
  url = {https://link.aps.org/doi/10.1103/PhysRevD.94.085012}
}

@article{ChangJH,
  title = {Supernova 1987A constraints on sub-GeV dark sectors, millicharged particles, the QCD axion, and an axion-like particle},
  author = {Chang, J. H. and Essig, R. and McDermott, S. D.},
  journal = {J. High Energ. Phys.},
  volume = {2018},
  pages = {51},
  year = {2018},
  doi = {10.1007/JHEP09(2018)051}
}

@article{Carenza,
  title = {Improved axion emissivity from a supernova via nucleon-nucleon bremsstrahlung},
  author = {Carenza, Perluca and Fischer, Tobias and Giannotti, Maurizio and Guo, Gang and Martinez-Pinedo, Gabriel and Mirizzi, Alessandro},
  journal = {JCAP},
  volume = {10},
  pages = {16},
  year = {2019},
  doi = {10.1088/1475-7516/2019/10/016}
}

@article{PhysRevLett.126.071102,
  title = {Enhanced Supernova Axion Emission and Its Implications},
  author = {Carenza, Pierluca and Fore, Bryce and Giannotti, Maurizio and Mirizzi, Alessandro and Reddy, Sanjay},
  journal = {Phys. Rev. Lett.},
  volume = {126},
  issue = {7},
  pages = {071102},
  numpages = {6},
  year = {2021},
  month = {Feb},
  publisher = {American Physical Society},
  doi = {10.1103/PhysRevLett.126.071102},
  url = {https://link.aps.org/doi/10.1103/PhysRevLett.126.071102}
}

@article{PhysRevLett.106.081101,
  title = {Rapid Cooling of the Neutron Star in Cassiopeia A Triggered by Neutron Superfluidity in Dense Matter},
  author = {Page, Dany and Prakash, Madappa and Lattimer, James M. and Steiner, Andrew W.},
  journal = {Phys. Rev. Lett.},
  volume = {106},
  issue = {8},
  pages = {081101},
  numpages = {4},
  year = {2011},
  month = {Feb},
  publisher = {American Physical Society},
  doi = {10.1103/PhysRevLett.106.081101},
  url = {https://link.aps.org/doi/10.1103/PhysRevLett.106.081101}
}

@article{10.1111/j.1745-3933.2011.01015.x,
    author = {Shternin, Peter S. and Yakovlev, Dmitry G. and Heinke, Craig O. and Ho, Wynn C. G. and Patnaude, Daniel J.},
    title = {Cooling neutron star in the Cassiopeia A supernova remnant: evidence for superfluidity in the core},
    journal = {Monthly Notices of the Royal Astronomical Society: Letters},
    volume = {412},
    number = {1},
    pages = {L108-L112},
    year = {2011},
    month = {03},
    issn = {1745-3925},
    doi = {10.1111/j.1745-3933.2011.01015.x},
    url = {https://doi.org/10.1111/j.1745-3933.2011.01015.x},
   }

@article{LEINSON201587,
title = {Superfluid phases of triplet pairing and rapid cooling of the neutron star in Cassiopeia A},
journal = {Physics Letters B},
volume = {741},
pages = {87-91},
year = {2015},
issn = {0370-2693},
doi = {https://doi.org/10.1016/j.physletb.2014.12.017},
url = {https://www.sciencedirect.com/science/article/pii/S0370269314008880},
author = {Lev B. Leinson}
}

@article{Leinson_2014,
doi = {10.1088/1475-7516/2014/08/031},
url = {https://dx.doi.org/10.1088/1475-7516/2014/08/031},
year = {2014},
month = {aug},
publisher = {},
volume = {2014},
number = {08},
pages = {031},
author = {Leinson, Lev B.},
title = {Axion mass limit from observations of the neutron star in Cassiopeia A},
journal = {Journal of Cosmology and Astroparticle Physics}
}

@article{PhysRevD.98.103015,
  title = {Limit on the axion decay constant from the cooling neutron star in Cassiopeia A},
  author = {Hamaguchi, Koichi and Nagata, Natsumi and Yanagi, Keisuke and Zheng, Jiaming},
  journal = {Phys. Rev. D},
  volume = {98},
  issue = {10},
  pages = {103015},
  numpages = {7},
  year = {2018},
  month = {Nov},
  publisher = {American Physical Society},
  doi = {10.1103/PhysRevD.98.103015},
  url = {https://link.aps.org/doi/10.1103/PhysRevD.98.103015}
}

@article{Leinson,
  title = {Impact of axions on the Cassiopea A neutron star cooling},
  author = {Leinson, Lev B.},
  journal = {JCAP},
  volume = {09},
  pages = {001},
  year = {2021},
  doi = {10.1088/1475-7516/2021/09/001}
}

@article{PhysRevD.93.065044,
  title = {Axion cooling of neutron stars},
  author = {Sedrakian, Armen},
  journal = {Phys. Rev. D},
  volume = {93},
  issue = {6},
  pages = {065044},
  numpages = {10},
  year = {2016},
  month = {Mar},
  publisher = {American Physical Society},
  doi = {10.1103/PhysRevD.93.065044},
  url = {https://link.aps.org/doi/10.1103/PhysRevD.93.065044}
}

@article{PhysRevD.99.043011,
  title = {Axion cooling of neutron stars. II. Beyond hadronic axions},
  author = {Sedrakian, Armen},
  journal = {Phys. Rev. D},
  volume = {99},
  issue = {4},
  pages = {043011},
  numpages = {10},
  year = {2019},
  month = {Feb},
  publisher = {American Physical Society},
  doi = {10.1103/PhysRevD.99.043011},
  url = {https://link.aps.org/doi/10.1103/PhysRevD.99.043011}
}

@article{RevModPhys.93.015004,
  title = {Invisible axion search methods},
  author = {Sikivie, Pierre},
  journal = {Rev. Mod. Phys.},
  volume = {93},
  issue = {1},
  pages = {015004},
  numpages = {36},
  year = {2021},
  month = {Feb},
  publisher = {American Physical Society},
  doi = {10.1103/RevModPhys.93.015004},
  url = {https://link.aps.org/doi/10.1103/RevModPhys.93.015004}
}

@article{PhysRevLett.51.1415,
  title = {Experimental Tests of the "Invisible" Axion},
  author = {Sikivie, P.},
  journal = {Phys. Rev. Lett.},
  volume = {51},
  issue = {16},
  pages = {1415--1417},
  numpages = {0},
  year = {1983},
  month = {Oct},
  publisher = {American Physical Society},
  doi = {10.1103/PhysRevLett.51.1415},
  url = {https://link.aps.org/doi/10.1103/PhysRevLett.51.1415}
}

@article{PhysRevD.64.092003,
  title = {Large-scale microwave cavity search for dark-matter axions},
  author = {Asztalos, S. and others},
  journal = {Phys. Rev. D},
  volume = {64},
  issue = {9},
  pages = {092003},
  numpages = {28},
  year = {2001},
  month = {Oct},
  publisher = {American Physical Society},
  doi = {10.1103/PhysRevD.64.092003},
  url = {https://link.aps.org/doi/10.1103/PhysRevD.64.092003}
}

@article{PhysRevLett.120.151301,
  title = {Search for Invisible Axion Dark Matter with the Axion Dark Matter Experiment},
  author = {Du, N. and others},
  collaboration = {ADMX Collaboration},
  journal = {Phys. Rev. Lett.},
  volume = {120},
  issue = {15},
  pages = {151301},
  numpages = {5},
  year = {2018},
  month = {Apr},
  publisher = {American Physical Society},
  doi = {10.1103/PhysRevLett.120.151301},
  url = {https://link.aps.org/doi/10.1103/PhysRevLett.120.151301}
}

@article{PhysRevLett.124.101303,
  title = {Extended Search for the Invisible Axion with the Axion Dark Matter Experiment},
  author = {Braine, T. and others},
  collaboration = {ADMX Collaboration},
  journal = {Phys. Rev. Lett.},
  volume = {124},
  issue = {10},
  pages = {101303},
  numpages = {6},
  year = {2020},
  month = {Mar},
  publisher = {American Physical Society},
  doi = {10.1103/PhysRevLett.124.101303},
  url = {https://link.aps.org/doi/10.1103/PhysRevLett.124.101303}
}

@article{PhysRevLett.127.261803,
  title = {Search for Invisible Axion Dark Matter in the $3.3--4.2\text{ }\text{ }\ensuremath{\mu}\mathrm{eV}$ Mass Range},
  author = {Bartram, C. and others},
  collaboration = {ADMX Collaboration},
  journal = {Phys. Rev. Lett.},
  volume = {127},
  issue = {26},
  pages = {261803},
  numpages = {6},
  year = {2021},
  month = {Dec},
  publisher = {American Physical Society},
  doi = {10.1103/PhysRevLett.127.261803},
  url = {https://link.aps.org/doi/10.1103/PhysRevLett.127.261803}
}

@article{PhysRevD.97.092001,
  title = {Results from phase 1 of the HAYSTAC microwave cavity axion experiment},
  author = {Zhong, L. and others},
  journal = {Phys. Rev. D},
  volume = {97},
  issue = {9},
  pages = {092001},
  numpages = {7},
  year = {2018},
  month = {May},
  publisher = {American Physical Society},
  doi = {10.1103/PhysRevD.97.092001},
  url = {https://link.aps.org/doi/10.1103/PhysRevD.97.092001}
}

@article{PhysRevD.106.103008,
  title = {Projected sensitivity of ${\text{DMRadio-m}}^{3}$: A search for the QCD axion below $1\text{ }\text{ }\mathrm{\ensuremath{\mu}}\mathrm{eV}$},
  author = {Brouwer, L. and others},
  collaboration = {DMRadio Collaboration},
  journal = {Phys. Rev. D},
  volume = {106},
  issue = {10},
  pages = {103008},
  numpages = {9},
  year = {2022},
  month = {Nov},
  publisher = {American Physical Society},
  doi = {10.1103/PhysRevD.106.103008},
  url = {https://link.aps.org/doi/10.1103/PhysRevD.106.103008}
}

@article{PhysRevLett.127.081801,
  title = {Search for Low-Mass Axion Dark Matter with ABRACADABRA-10 cm},
  author = {Salemi, Chiara P. and others},
  journal = {Phys. Rev. Lett.},
  volume = {127},
  issue = {8},
  pages = {081801},
  numpages = {7},
  year = {2021},
  month = {Aug},
  publisher = {American Physical Society},
  doi = {10.1103/PhysRevLett.127.081801},
  url = {https://link.aps.org/doi/10.1103/PhysRevLett.127.081801}
}

@article{PhysRevX.4.021030,
  title = {Proposal for a Cosmic Axion Spin Precession Experiment (CASPEr)},
  author = {Budker, Dmitry and Graham, Peter W. and Ledbetter, Micah and Rajendran, Surjeet and Sushkov, Alexander O.},
  journal = {Phys. Rev. X},
  volume = {4},
  issue = {2},
  pages = {021030},
  numpages = {10},
  year = {2014},
  month = {May},
  publisher = {American Physical Society},
  doi = {10.1103/PhysRevX.4.021030},
  url = {https://link.aps.org/doi/10.1103/PhysRevX.4.021030}
}

@misc{quiskamp2024nearquantumlimitedaxiondark,
      title={Near-quantum limited axion dark matter search with the ORGAN experiment around 26 $\mu$eV}, 
      author={Quiskamp, A. P. and others},
      year={2024},
      eprint={2407.18586},
      archivePrefix={arXiv},
      primaryClass={hep-ex},
      url={https://arxiv.org/abs/2407.18586}
}

@article{CAST,
title = {New CAST limit on the axion-photon interaction},
author = {Anastassopoulos, V. and others},
collaboration= {CAST Collaboration},
journal = {Nature Physics},
volume = {13},
pages = {584-590},
year = {2017},
doi = {https://doi.org/10.1038/nphys4109}
}

@article{PhysRevLett.66.2557,
  title = {Red-giant evolution, metallicity, and new bounds on hadronic axions},
  author = {Haxton, W. C. and Lee, K. Y.},
  journal = {Phys. Rev. Lett.},
  volume = {66},
  issue = {20},
  pages = {2557--2560},
  numpages = {0},
  year = {1991},
  doi = {10.1103/PhysRevLett.66.2557},
  url = {https://link.aps.org/doi/10.1103/PhysRevLett.66.2557}
}

@article{PhysRevLett.75.3222,
  title = {Proposal to Search for a Monochromatic Component of Solar Axions Using ${}^{57}$Fe},
  author = {Moriyama, Shigetaka},
  journal = {Phys. Rev. Lett.},
  volume = {75},
  issue = {18},
  pages = {3222--3225},
  numpages = {0},
  year = {1995},
  month = {Oct},
  publisher = {American Physical Society},
  doi = {10.1103/PhysRevLett.75.3222},
  url = {https://link.aps.org/doi/10.1103/PhysRevLett.75.3222}
}

@article{CASTcollaboration_2009,
doi = {10.1088/1475-7516/2009/12/002},
url = {https://dx.doi.org/10.1088/1475-7516/2009/12/002},
year = {2009},
month = {dec},
publisher = {},
volume = {2009},
number = {12},
pages = {002},
collaboration = {CAST Collaboration},
author = {Andriamonje, S. and others},
title = {Search for 14.4 keV solar axions emitted in the M1-transition of 57Fe nuclei with CAST},
journal = {Journal of Cosmology and Astroparticle Physics}
}

@article{KRCMAR199838,
title = {Search for solar axions using 57Fe},
journal = {Phys. Letts. B},
author = {Krcmar, M. and Krecak, Z. and Stipcevic, M. and Ljubicic, A. and Bradley, D. A.},
volume = {442},
pages = {38-42},
year = {1998},
issn = {0370-2693},
doi = {https://doi.org/10.1016/S0370-2693(98)01231-3},
url = {https://www.sciencedirect.com/science/article/pii/S0370269398012313}
}

@article{Derbin,
title = {New limit on the mass of 14.4-keV solar axions emitted in an M1 transition in 57Fe nuclei},
author = {Derbin, A. V. and Muratova, V. N. and Semenov, D. A. and Unzhakov, E. V.},
journal = {Physics of Atomic Nuclei},
volume = {74},
pages = {596-602},
year = {2011},
doi = {https://doi.org/10.1134/S1063778811040041}
}

@article{NAMBA2007398,
title = {Results of a search for monochromatic solar axions using 57Fe},
journal = {Physics Letters B},
volume = {645},
number = {5},
pages = {398-401},
year = {2007},
issn = {0370-2693},
doi = {https://doi.org/10.1016/j.physletb.2007.01.005},
url = {https://www.sciencedirect.com/science/article/pii/S0370269307000615},
author = {T. Namba}
}

@article{Cortona,
title = {The QCD axion,precisely},
author = {di Cortona, G. G. and Hardy, E. and Vega, J. P.},
journal = {J. High Energy Physics},
volume = {2016},
pages = {34},
year = {2016},
doi = {https://doi.org/10.1007/JHEP01(2016)034}
}

@article{PhysRevLett.43.103,
  title = {Weak-Interaction Singlet and Strong $\mathrm{CP}$ Invariance},
  author = {Kim, Jihn E.},
  journal = {Phys. Rev. Lett.},
  volume = {43},
  issue = {2},
  pages = {103--107},
  numpages = {0},
  year = {1979},
  month = {Jul},
  publisher = {American Physical Society},
  doi = {10.1103/PhysRevLett.43.103},
  url = {https://link.aps.org/doi/10.1103/PhysRevLett.43.103}
}

@article{SHIFMAN1980493,
title = {Can confinement ensure natural CP invariance of strong interactions?},
journal = {Nuclear Physics B},
volume = {166},
number = {3},
pages = {493-506},
year = {1980},
issn = {0550-3213},
doi = {https://doi.org/10.1016/0550-3213(80)90209-6},
url = {https://www.sciencedirect.com/science/article/pii/0550321380902096},
author = {M.A. Shifman and A.I. Vainshtein and V.I. Zakharov}
}

@article{DINE1981199,
title = {A simple solution to the strong CP problem with a harmless axion},
journal = {Physics Letters B},
volume = {104},
number = {3},
pages = {199-202},
year = {1981},
issn = {0370-2693},
doi = {https://doi.org/10.1016/0370-2693(81)90590-6},
url = {https://www.sciencedirect.com/science/article/pii/0370269381905906},
author = {Michael Dine and Willy Fischler and Mark Srednicki}
}

@article{osti_7063072,
  author       = {Zhitnitskii, A P},
  title        = {Possible suppression of axion-hadron interactions},
  url          = {https://www.osti.gov/biblio/7063072},
  journal      = {Sov. J. Nucl. Phys. (Engl. Transl.); (United States)},
  volume       = {31:2},
  place        = {United States},
  year         = {1980},
  month        = {02}
  }

@article{Limongi_2024,
doi = {10.3847/1538-4365/ad12c1},
url = {https://dx.doi.org/10.3847/1538-4365/ad12c1},
year = {2024},
month = {jan},
publisher = {The American Astronomical Society},
volume = {270},
number = {2},
pages = {29},
author = {Limongi, Marco and Roberti, Lorenzo and Chieffi, Alessandro and Nomoto, Ken’ichi},
title = {Evolution and Final Fate of Solar Metallicity Stars in the Mass Range 7-15 Solar $M_\odot$. I. The Transition from Asymptotic Giant Branch to Super-AGB Stars, Electron Capture, and Core-collapse Supernova Progenitors},
journal = {The Astrophysical Journal Supplement Series}
}

@ARTICLE{Paxton2011,
  author = {{Paxton}, B. and {Bildsten}, L. and {Dotter}, A. and {Herwig}, F. and {Lesaffre}, P. and {Timmes}, F.},
  title = {{Modules for Experiments in Stellar Astrophysics (MESA)}},
  journal = {Astrophys. J. Suppl.},
  archivePrefix = {arXiv},
  eprint = {1009.1622},
  primaryClass = {astro-ph.SR},
  year = {2011},
  month = {jan},
  volume = {192},
  eid = {3},
  pages = {3},
  doi = {10.1088/0067-0049/192/1/3},
  adsurl = {https://ui.adsabs.harvard.edu/abs/2011ApJS..192....3P}
}

@ARTICLE{Paxton2013,
  author = {Paxton, B. and others},
  title = {{Modules for Experiments in Stellar Astrophysics (MESA): Planets, Oscillations, Rotation, and Massive Stars}},
  journal = {Astrophys. J. Suppl.},
  archivePrefix = {arXiv},
  eprint = {1301.0319},
  primaryClass = {astro-ph.SR},
  year = {2013},
  month = {sep},
  volume = {208},
  eid = {4},
  pages = {4},
  doi = {10.1088/0067-0049/208/1/4},
  adsurl = {https://ui.adsabs.harvard.edu/abs/2013ApJS..208....4P}
}

@ARTICLE{Paxton2015,
  author = {Paxton, B. and others},
  title = {{Modules for Experiments in Stellar Astrophysics (MESA): Binaries, Pulsations, and Explosions}},
  journal = {Astrophys. J. Suppl.},
  archivePrefix = {arXiv},
  eprint = {1506.03146},
  primaryClass = {astro-ph.SR},
  year = {2015},
  month = {sep},
  volume = {220},
  eid = {15},
  pages = {15},
  doi = {10.1088/0067-0049/220/1/15},
  adsurl = {https://ui.adsabs.harvard.edu/abs/2015ApJS..220...15P}
}

@ARTICLE{Paxton2018,
  author = {Paxton, B. and others},
  title = {{Modules for Experiments in Stellar Astrophysics (MESA): Convective Boundaries, Element Diffusion, and Massive Star Explosions}},
  journal = {Astrophys. J. Suppl.},
  archivePrefix = {arXiv},
  eprint = {1710.08424},
  primaryClass = {astro-ph.SR},
  year = {2018},
  month = {feb},
  volume = {234},
  eid = {34},
  pages = {34},
  doi = {10.3847/1538-4365/aaa5a8},
  adsurl = {https://ui.adsabs.harvard.edu/abs/2018ApJS..234...34P}
}

@ARTICLE{Paxton2019,
       author = {{Paxton}, B. and others},
        title = "{Modules for Experiments in Stellar Astrophysics (MESA): Pulsating Variable Stars, Rotation, Convective Boundaries, and Energy Conservation}",
      journal = {Astrophys. J. Suppl.},
         year = {2019},
       volume = {243},
       number = {1},
          eid = {10},
        pages = {10},
          doi = {10.3847/1538-4365/ab2241},
archivePrefix = {arXiv},
       eprint = {1903.01426},
 primaryClass = {astro-ph.SR},
       adsurl = {https://ui.adsabs.harvard.edu/abs/2019ApJS..243...10P}
}

@ARTICLE{Jermyn2023,
       author = {{Jermyn}, A. S. and others},
        title = "{Modules for Experiments in Stellar Astrophysics (MESA): Time-dependent Convection, Energy Conservation, Automatic Differentiation, and Infrastructure}",
      journal = {Astrophys. J. Suppl.},
         year = 2023,
        month = mar,
       volume = {265},
       number = {1},
          eid = {15},
        pages = {15},
          doi = {10.3847/1538-4365/acae8d},
archivePrefix = {arXiv},
       eprint = {2208.03651},
 primaryClass = {astro-ph.SR},
       adsurl = {https://ui.adsabs.harvard.edu/abs/2023ApJS..265...15J}
}

@article{SHAMSUZZOHABASUNIA20211,
title = {Nuclear Data Sheets for A=23},
journal = {Nuclear Data Sheets},
volume = {171},
pages = {1-252},
year = {2021},
issn = {0090-3752},
doi = {https://doi.org/10.1016/j.nds.2020.12.001},
url = {https://www.sciencedirect.com/science/article/pii/S0090375220300582},
author = {{Shamsuzzoha Basunia}, M. and Chakraborty, A.}
}

@article{Verberne:2021tse,
	author = "Verberne, Sill and Vink, Jacco",
	title = "{The radial supernova remnant distribution in the Galaxy}",
	eprint = "2103.16973",
	archivePrefix = "arXiv",
	primaryClass = "astro-ph.GA",
	doi = "10.1093/mnras/stab940",
	journal = "Mon. Not. Roy. Astron. Soc.",
	volume = "504",
	number = "1",
	pages = "1536--1544",
	year = "2021"
}

@ARTICLE{1977ApJ...217..843S,
       author = {{Stecker}, F.~W. and {Jones}, F.~C.},
        title = "{The galactic halo question: new size constraints from galactic gamma -ray data.}",
      journal = {Astrophys. J.},
         year = 1977,
        month = nov,
       volume = {217},
        pages = {843-851},
          doi = {10.1086/155631},
       adsurl = {https://ui.adsabs.harvard.edu/abs/1977ApJ...217..843S}
}

@article{Case_1998,
doi = {10.1086/306089},
url = {https://dx.doi.org/10.1086/306089},
year = {1998},
month = {sep},
publisher = {},
volume = {504},
number = {2},
pages = {761},
author = {Case, Gary L. and Bhattacharya, Dipen},
title = {A New $\Sigma$-D Relation and Its Application to the Galactic Supernova Remnant Distribution},
journal = {Astrophys. J.}
}

@ARTICLE{1955ApJ...121..161S,
       author = {{Salpeter}, Edwin E.},
        title = "{The Luminosity Function and Stellar Evolution.}",
      journal = {Astrophys. J.},
         year = 1955,
        month = jan,
       volume = {121},
        pages = {161},
          doi = {10.1086/145971},
       adsurl = {https://ui.adsabs.harvard.edu/abs/1955ApJ...121..161S}
}

@article{Williams_2018,
doi = {10.3847/1538-4357/aaba7d},
url = {https://dx.doi.org/10.3847/1538-4357/aaba7d},
year = {2018},
month = {jun},
publisher = {The American Astronomical Society},
volume = {860},
number = {1},
pages = {39},
author = {Williams, B. F. and Hillis, T. J. and Murphy, J. W. and Gilbert, K. and Dalcanton, J. J. and Dolphin, A. E.},
title = {Constraints for the Progenitor Masses of Historic Core-collapse Supernovae},
journal = {Astrophys. J.}
}

@article{ROZWADOWSKA2021101498,
title = {On the rate of core collapse supernovae in the milky way},
author = {Rozwadowska, K. and Vissani, F. and Cappellaro, E.},
journal = {New Astronomy},
volume = {83},
pages = {101498},
year = {2021},
issn = {1384-1076},
doi = {https://doi.org/10.1016/j.newast.2020.101498},
url = {https://www.sciencedirect.com/science/article/pii/S138410762030289X}
}

@article{Planck,
title = {Planck 2018 results: VI. Cosmological parameters},
author = {Aghanim, N. and others},
collaboration = {Planck Collaboration},
journal = {A\&A},
volume = {641},
pages = {A6},
year = {2020},
doi = {https://doi.org/10.1038/nature04364}
}

@article{Diehl,
title = {Radioactive $^{26}$Al from massive stars in the Galaxy},
author = {Diehl, R. and others},
journal = {Nature},
volume = {439},
pages = {45},
year = {2006},
doi = {https://doi.org/10.1038/nature04364}
}

@article{10.1046/j.1365-8711.1999.02145.x,
    author = {Dragicevich, P. M. and Blair, D. G. and Burman, R. R.},
    title = {Why are supernovae in our Galaxy so frequent?},
    journal = {Monthly Notices of the Royal Astronomical Society},
    volume = {302},
    number = {4},
    pages = {693-699},
    year = {1999},
    month = {02},
    doi = {10.1046/j.1365-8711.1999.02145.x},
    url = {https://doi.org/10.1046/j.1365-8711.1999.02145.x}
}

@article{KM,
title = {Galactic distribution of supernovae and OB associations},
author = {Kachelriess, M. and Mikalsen, V.},
journal = {Comp. Phys. Comm.},
volume = {311},
pages = {109537},
year = {2025},
doi = {https://doi.org/10.1016/j.cpc.2025.109537}
}

@article{Camisassa,
title = {The evolution of ultra-masssive white dwarfs},
author = {Camisassa, M. E. and others},
journal = {A\&A},
volume = {625},
pages = {A87},
year = {2019},
doi = {A&A 625, A87 (2019)
https://doi.org/10.1051/0004-6361/201833822}
}

@article{Cunningham,
    author = {Cunningham, T. and Tremblay, P.-E. and O'Brien, M. W.},
    title = {Initial-final mass relation from white dwarfs within 40?pc},
    journal = {MNRAS},
    volume = {527},
    number = {2},
    pages = {3602},
    year = {2023},
    doi = {10.1093/mnras/stad3275}
}

@article{LI2024139075,
title = {Upper limit on the axion-photon coupling from Markarian 421},
journal = {Physics Letters B},
volume = {858},
pages = {139075},
year = {2024},
issn = {0370-2693},
doi = {https://doi.org/10.1016/j.physletb.2024.139075},
url = {https://www.sciencedirect.com/science/article/pii/S0370269324006336},
author = {Hai-Jun Li and Wei Chao and Yu-Feng Zhou},
}

@article{Unger_2024,
doi = {10.3847/1538-4357/ad4a54},
url = {https://dx.doi.org/10.3847/1538-4357/ad4a54},
year = {2024},
month = {jul},
publisher = {The American Astronomical Society},
volume = {970},
number = {1},
pages = {95},
author = {Unger, Michael and Farrar, Glennys R.},
title = {The Coherent Magnetic Field of the Milky Way},
journal = {The Astrophysical Journal}
}

@article{PhysRevD.37.1237,
	title = {Mixing of the photon with low-mass particles},
	author = {Raffelt, Georg and Stodolsky, Leo},
	journal = {Phys. Rev. D},
	volume = {37},
	issue = {5},
	pages = {1237--1249},
	numpages = {0},
	year = {1988},
	month = {Mar},
	publisher = {American Physical Society},
	doi = {10.1103/PhysRevD.37.1237},
	url = {https://link.aps.org/doi/10.1103/PhysRevD.37.1237}
}

@article{CUORE:2012ymr,
    author = "Alessandria, F. and others",
    collaboration = "CUORE",
    title = "{Search for 14.4 keV solar axions from M1 transition of Fe-57 with CUORE crystals}",
    eprint = "1209.2800",
    archivePrefix = "arXiv",
    primaryClass = "hep-ex",
    doi = "10.1088/1475-7516/2013/05/007",
    journal = "JCAP",
    volume = "05",
    pages = "007",
    year = "2013"
}

@article{DiLuzio:2021qct,
    author = "Di Luzio, Luca and others",
    title = "{Probing the axion\textendash{}nucleon coupling with the next generation of~axion helioscopes}",
    eprint = "2111.06407",
    archivePrefix = "arXiv",
    primaryClass = "hep-ph",
    reportNumber = "DESY-21-194",
    doi = "10.1140/epjc/s10052-022-10061-1",
    journal = "Eur. Phys. J. C",
    volume = "82",
    number = "2",
    pages = "120",
    year = "2022"
}

@article{Fleury:2022plh,
    author = "Fleury, Leesa and Caiazzo, Ilaria and Heyl, Jeremy",
    title = "{Constraining axions with ZTF J1901+1458}",
    eprint = "2208.00405",
    archivePrefix = "arXiv",
    primaryClass = "astro-ph.HE",
    doi = "10.1103/PhysRevD.107.L101303",
    journal = "Phys. Rev. D",
    volume = "107",
    number = "10",
    pages = "L101303",
    year = "2023"
}

@article{Carenza:2024ehj,
    author = "Carenza, Pierluca and Giannotti, Maurizio and Isern, Jordi and Mirizzi, Alessandro and Straniero, Oscar",
    title = "{Axion astrophysics}",
    eprint = "2411.02492",
    archivePrefix = "arXiv",
    primaryClass = "hep-ph",
    reportNumber = "BARI-TH/66-24",
    doi = "10.1016/j.physrep.2025.02.002",
    journal = "Phys. Rept.",
    volume = "1117",
    pages = "1--102",
    year = "2025"
}

@article{Tomsick:2023aue,
    author = "Tomsick, John A. and others",
    title = "{The Compton Spectrometer and Imager}",
    eprint = "2308.12362",
    archivePrefix = "arXiv",
    primaryClass = "astro-ph.HE",
    doi = "10.22323/1.444.0745",
    journal = "PoS",
    volume = "ICRC2023",
    pages = "745",
    year = "2023"
}

@article{PhysRevLett.133.211002,
  title = {Supernova Axions Convert to Gamma Rays in Magnetic Fields of Progenitor Stars},
  author = {Manzari, Claudio Andrea and Park, Yujin and Safdi, Benjamin R. and Savoray, Inbar},
  journal = {Phys. Rev. Lett.},
  volume = {133},
  issue = {21},
  pages = {211002},
  numpages = {10},
  year = {2024},
  month = {Nov},
  publisher = {American Physical Society},
  doi = {10.1103/PhysRevLett.133.211002},
  url = {https://link.aps.org/doi/10.1103/PhysRevLett.133.211002}
}

@article{Benabou:2025jcv,
    author = "Benabou, Joshua N. and Dessert, Christopher and Patra, Kishore C. and Brink, Thomas G. and Zheng, WeiKang and Filippenko, Alexei V. and Safdi, Benjamin R.",
    title = "{Search for Axions in Magnetic White Dwarf Polarization at Lick and Keck Observatories}",
    eprint = "2504.12377",
    archivePrefix = "arXiv",
    primaryClass = "hep-ph",
    month = "4",
    year = "2025"
}

@article{Ning:2024eky,
    author = "Ning, Orion and Safdi, Benjamin R.",
    title = "{Leading Axion-Photon Sensitivity with NuSTAR Observations of M82 and M87}",
    eprint = "2404.14476",
    archivePrefix = "arXiv",
    primaryClass = "hep-ph",
    doi = "10.1103/PhysRevLett.134.171003",
    journal = "Phys. Rev. Lett.",
    volume = "134",
    number = "17",
    pages = "171003",
    year = "2025"
}

@article{Buschmann:2021juv,
    author = "Buschmann, Malte and Dessert, Christopher and Foster, Joshua W. and Long, Andrew J. and Safdi, Benjamin R.",
    title = "{Upper Limit on the QCD Axion Mass from Isolated Neutron Star Cooling}",
    eprint = "2111.09892",
    archivePrefix = "arXiv",
    primaryClass = "hep-ph",
    doi = "10.1103/PhysRevLett.128.091102",
    journal = "Phys. Rev. Lett.",
    volume = "128",
    number = "9",
    pages = "091102",
    year = "2022"
}

@misc{GitHub,
  title        = {Stellar axion simulation},
  author       = {Liu, Xing},
  year         = 2025,
  note         = {\url{https://github.com/xingyzt/saltyaxions}}
}

\end{document}